\documentclass[useAMS,usenatbib]{mn2e}
\usepackage{graphicx,amssym}
\newcommand{\bc}{\begin{center}}
\newcommand{\ec}{\end{center}}

\title[ Are mergers responsible for universal halo properties?]
      { Are mergers responsible for universal halo properties?}
\author[J. ~Wang \& S. D. M. White]
       {Jie Wang\thanks{Email: wangjie@mpa-garching.mpg.de},
       Simon D.~M.~White\thanks{Email: swhite@mpa-garching.mpg.de}
        \\
        Max--Planck--Institut f\"ur Astrophysik,
        Karl--Schwarzschild--Str. 1, D-85748 Garching, Germany}
\begin{document}

\date{Accepted 2008 ???? ??.
      Received 2008 ???? ??;
      in original form 2008 ???? ??}

\pagerange{\pageref{firstpage}--\pageref{lastpage}}
\pubyear{2007}

\maketitle

\label{firstpage}

\begin{abstract}
 $N$-body simulations of Cold Dark Matter (CDM) have shown that, in
this hierarchical structure formation model, dark matter halo
properties, such as the density profile, the phase-space density
profile, the distribution of axial ratio, the distribution of spin
parameter, and the distribution of internal specific angular
momentum follow `universal' laws or distributions. Here we study the
properties of the first generation of haloes in a Hot Dark Matter
(HDM) dominated universe, as an example of halo formation through
monolithic collapse. We find all these universalities to be present
in this case also. Halo density profiles are very well fit by the
Navarro et al (1997) profile over two orders of magnitude in mass.
The concentration parameter depends on mass as $c \propto M^{0.2}$,
reversing the dependence found in a hierarchical CDM universe.
However, the concentration-formation time relation is similar in the
two cases: earlier forming haloes tend to be more concentrated than
their later forming counterparts. Halo formation histories are also
characterized by two phases in the HDM case: an early phase of rapid
accretion followed by slower growth. Furthermore, there is no
significant difference between the HDM and CDM cases concerning the
statistics of other halo properties: the phase-space density
profile; the velocity anisotropy profile; the distribution of
shape parameters; the distribution of spin parameter, and the
distribution of internal specific angular momentum are all similar
in the two cases. Only substructure content differs dramatically. 
These results indicate that mergers do not play a
pivotal role in establishing the universalities, thus
contradicting models which explain them as consequences of mergers.
\end{abstract}


\section{Introduction}
\label{sec:intro}

The mass distribution of the self-gravitating, quasi-equilibrium
dark haloes that form in an expanding universe is an issue of
fundamental importance. Early work on self-similar spherical
collapse predicts virialized structures with power-law density
profiles \citep{ fillmore84,bertschinger85}. As $N$-body techniques
improved, it was realised that, in the hierarchical universes, the
profile departs significantly from a single power law and is better
fitted by a profile with curvature in a log-log plot.
\citep{efstathiou85,dubinski91}. Later, it was established that the
density profiles of haloes in CDM and other hierarchical clustering
cosmologies, have a universal form which is well represented by the
simple fitting formula of \citet[][hereafter NFW]{nfw96,nfw97}
\begin{equation}
\rho(r)=\frac{\rho_s}{(r/r_s)(1+r/r_s)^2}
\end{equation}
where $r_s$ is a characteristic radius where the logarithmic density
slope is $-2$, and $\rho_s/4$ is the density at $r_s$. The above
equation implies that $\rho\propto r^{-1}$ in the inner regions and
$\rho \propto r^{-3}$ in the outskirts. A useful alternative
parameter for describing the shape of the profile is the
concentration parameter $c=r_{200}/r_s$ ($r_{200}$ is the virial
radius  defined as the radius within which the mean density is
$200$ times the critical value). Shortly after the NFW papers,
\citet{huss99b} showed that the NFW model also fit halos which form
by monolithic collapse (i.e. without mergers) in a HDM cosmology.
Nevertheless much work has focused on the role of mergers 
in establishing the NFW profile, 
since mergers contribute substantially to the growth of haloes
in the CDM model \citep{raig98,salvador98,syer98,
subramanian00,dekel03}. \citet{syer98} analysed dynamical processes
during repeated mergers, concluding that the universal profile is
generated by tidal stripping of small haloes as they merge into
larger objects. \citet{dekel03} extended this model and found the
tidal compression from a halo core makes the satellite orbits
decay from the radius where $\rho \propto r$ to the halo center and
causes a rapid steepening of the inner profile to $\rho \propto
r^{-\alpha} (\alpha >1)$. Recent work has confirmed the NFW
hypothesis, extending the comparison to a much wider range of
densities and radii than the original work, and uncovering small but
significant deviation, between the mean density profiles of
simulated dark halos and the NFW formula
\citep{power03,diemand04,merritt06,graham06,gao08,hayashi08}. The
differences, however, are quite small compared to the scatter
between different haloes of the same mass.

In addition to the density profile, other halo properties are found
to follow universal profiles or universal distributions  in the
hierarchical CDM model. The ``phase-space density''
$\rho(r)/\sigma^2(r)$ has a remarkably accurate pure power-law
distribution with radius \citep{taylor01,barnes06}. The
distributions of shape parameters (e.g. the axis ratios) have a weak
dependence on mass and redshift \citep{bullock02b,kasun05,
allgood06,bett07}. The spin parameter distribution is well described
by a log-normal function which varies weakly with halo mass
\citep{barnes87,warren92, cole96, bullock02a,bett07}.
\citet{bullock01b} also found that the cumulative mass distribution
of specific angular momentum $j$ is well fitted by a universal
function. All these distributions appear to depend little, if at
all, on the global cosmological parameters and on the shape of the
initial matter power spectrum. However, the origin of these
universalities is still not well understood.

In this paper, we will test whether mergers are the dominant
physical mechanism to produce these universal profiles or
distributions. To this end, we study a range of properties of the
haloes which grow by two different paradigms, by  hierarchical
aggregation and monolithic collapse.  The concordance cosmology
($\Lambda$CDM) is  a standard hierarchical universe, and the 
HDM dominated universe provides an example
where formation of the first haloes is monolithic. In a HDM
universe, small objects cannot form because free-streaming effects
smooth out small-scale structure in the initial condition. The first
halos then form by smooth collapse.

We begin in section~\ref{sec:simulation} with a summary of the
$N$-body simulations and halo catalogues used in this paper. In
section~\ref{sec:haloform}, we give a general description of how
halos form and evolve in the HDM universe. In section
~\ref{sec:result}, a range of halo properties are studied in depth
in our two model universes. Then, we will summarise and discuss 
our results in section~\ref{sec:conclusions}.

\begin{figure}
\bc
\hspace{-1.4cm}
\resizebox{9cm}{!}{\includegraphics{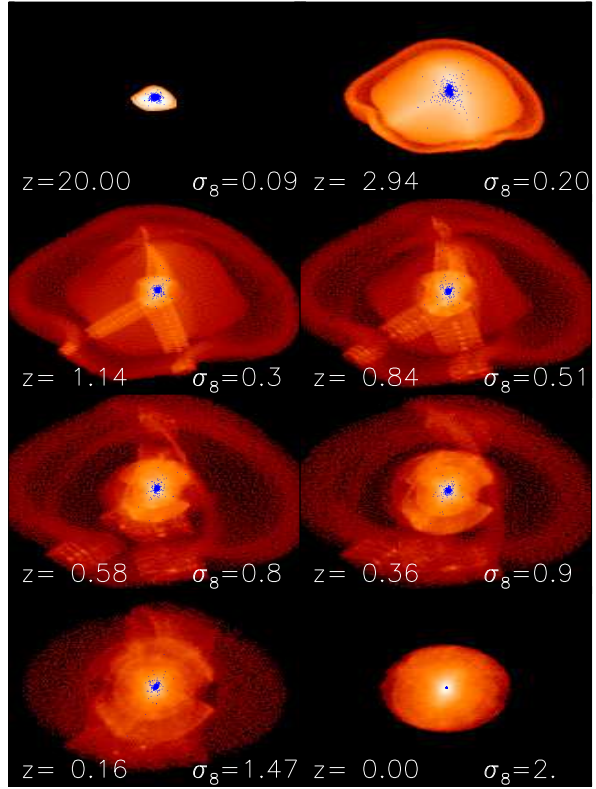}}\\%
\caption{The evolution of a massive halo in a HDM universe with mass
$1.3\times 10^{15} M_{\odot}/h$ at redshift zero. All particles
within $r_{200}$ at $z=0$ are traced back to seven higher
redshifts. In addition the particles in the very inner region
$r < 0.01 r_{200}$ are traced back and are shown as the blue points.
The size of each box is $\rm 8 Mpc/h$ in physical (not comoving)
units. The redshift and the corresponding $\sigma_8$ are
showed in each panel.}
\label{fig:halo_evo}
\ec
\end{figure}
\begin{figure}
\bc
\hspace{-1.4cm}
\resizebox{9cm}{!}{\includegraphics{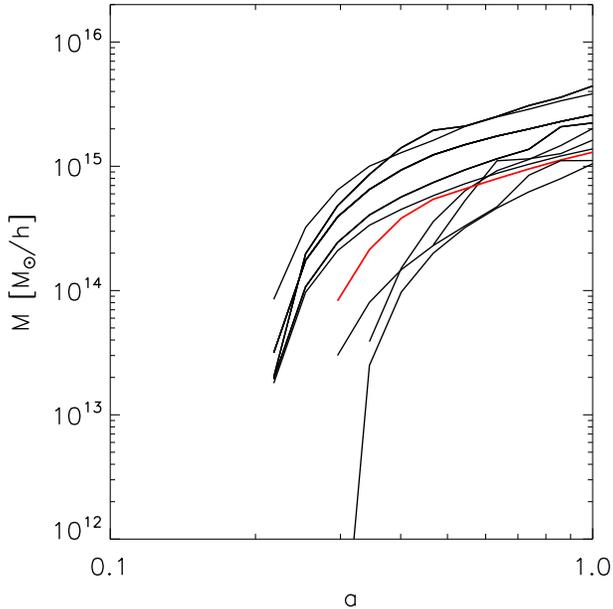}}\\%
\caption{Mass assembly histories (MAH) for the 10 most massive
haloes at $z=0$ in the HDMRUN sample. The red curve indicates the halo
presented in Fig.~\ref{fig:halo_evo}} \label{fig:mah} \ec
\end{figure}

\begin{figure}
\bc
\hspace{-1.4cm}
\resizebox{9cm}{!}{\includegraphics{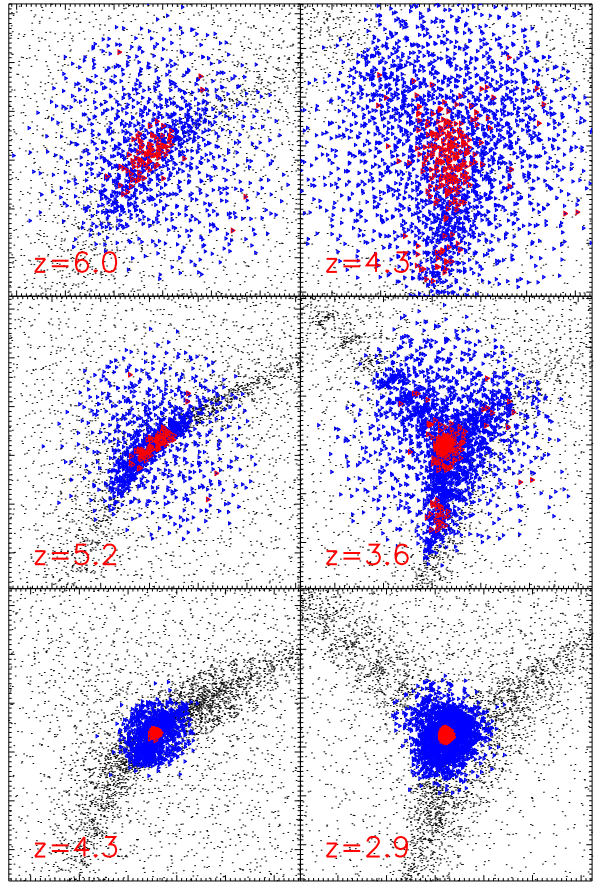}}\\%
\caption{Formation histories of two ``first'' objects in a HDM
universe. The right column shows a halo with $N_{200}=2686$ at
$z=2.9$ (roughly $M_{200}=3.5\times 10^{14}M_{\odot}$); while the
left column shows another halo with $N_{200}=1215$
($M_{200}=1.6\times 10^{14}M_{\odot}$) at $z=4.3$. All particles in
both halos at the last redshift are traced back to higher redshifts
and presented as blue points. Particles in the very inner region $r
< 0.1 r_{200}$  are traced back and presented as red points. Other
background particles are shown with black points. The coordinate is
comoving scale, and the size of each box is $\rm 6 Mpc/h$. The
redshift is labeled in each panel. The simulation glass128 (see
Paper I) is used here. Note the artificial discreteness features
visible in the filaments in the $z=3.6$ panel.}
\label{fig:halo_evo2} \ec
\end{figure}

\section{The simulations and halo catalogues}
\label{sec:simulation} In this work, we simulate the HDM density
field using $512^3$ particles within a $\rm 100 h^{-1}Mpc$ cube. For
simplicity, we choose an Einstein-de Sitter universe dominated by a
single massive neutrino. Then the cosmological matter density
parameter is $\Omega_m=1$ and the simulation particle mass is $2.07
\times 10^9 M_{\odot}/h$. The initial power spectrum used to perturb
the initial particle set is based on the theoretical prediction of
\citet{bond83}. The power spectrum in a HDM universe possesses a
coherent free streaming scale: $\rm \delta_{nu}=22.2 Mpc$ in our
case. The corresponding damping mass is $7.28 \times
10^{14}M_{\odot}/h$. This power spectrum is normalised to
$\sigma_8=2$ which corresponds to the collapse of the first
non-linear structures at $z \sim 6$. This simulation was
presented as glass512 in \citet[][ hereafter paper
I]{wang07a} and more technical details can be found there.  We
choose to compare it with a simulation of the WMAP1 model from
\citet{wang07b}. This simulates a concordance $\Lambda$CDM universe
with $540^3$ particles in a $\rm 125 h^{-1}Mpc$ cube. Hereafter we
refer these two simulations as HDMRUN and CDMRUN. The softening lengths
for these two simulations are $10$  and $\rm 5 kpc/h$ in the HDM and
CDM cases respectively.

As pointed out in Paper I, the population of haloes in HDMRUN is
contaminated because discreteness effects break up the filaments and
produce many artificial small haloes. The resolution limit above
which haloes are immune to this effect is $M_{lim}=8.8 \times
10^{12} M_{\odot}/h$, and corresponds to about 5000 particles. In
the current study we therefore focus primarily on well resolved
haloes with particle number $N_{200} > 10000$ within virial radius
$r_{200}$ in two simulations. Several of our most massive 
HDM haloes experience one or two major mergers at 
low redshift as in the CDM case. We exclude these second generation 
haloes from our analysis below.
The number of haloes which satisfy the above constraints is $58$ 
and $304$ in HDMRUN and CDMRUN respectively. When studying the $M_{200}$-c
relation, we reduce the particle number limit to 5000 in order to
cover a wider mass range. We then have $84$ (HDMRUN) and
$1752$ (CDMRUN) haloes in the two simulations.

Our halos are identified by a standard $b=0.2$ friends-of-friends
(FOF) group-finder \citep{davis85}. Then the SUBFIND algorithm
\citep{springel01} is used to resolve these objects into
substructures and main subhaloes. The latter then define our halo
sample through our $N_{200}$ limits. At high redshift, the FOF
method tends to link more particles together especially in our
HDMRUN where large filaments and sheets are easily joined. 
SUBFIND defines the centre as the minimum
of the gravitational potential, and this is used to estimate
$N_{200}$. We also use merger trees built up from the subhalo
catalogue produced by the SUBFIND algorithm \citep[see][]{springel05}.

\section{Monolithic Growth}
\label{sec:haloform} In Fig:~\ref{fig:halo_evo}, we study the
evolution of a typical halo in our HDMRUN simulation. The mass of
this halo is about twice the free-streaming mass scale. In the lower
right corner ($z=0$), all particles within $r_{200}$ are chosen.
These are then traced back to the initial condition in the other
panels. The particles within the central region $ r < 0.01 r_{200}$
are also traced back to high redshift and are highlighted with blue
points. It is interesting that these images are not similar to those
found in the concordance CDM universe using a similar representation
by \citet{gao04a}: the whole density field is smooth at all times
and no obvious substructures are seen. The collapse along the
filaments into knots occurs simultaneously with the accretion of
diffuse particles. At redshift $1.14$ and $0.84$, some parts of the
filaments are missing since they fall outside $r_{200}$ at $z=0$.
The particles which end up within $0.01 r_{200}$ stayed close
together at all times, even in the initial condition. They fall to
the centre in a smooth spherical collapse. In a CDM universe, as
presented in Fig 2. of \citet{gao04a}, the matter which ends up in
the central region typically comes from a number of different
objects at early times.

In Fig~\ref{fig:mah} we present the mass assembly history (MAH) of
the 10 most massive haloes in the HDMRUN sample. This is just the
mass growth of the most massive progenitor in each case. We find
that the MAHs are similar to  those of haloes in a CDM universe: at
early stages, the mass grows rapidly; but this slows
dramatically towards $z=0$. For example, the MAH of the halo shown
in Fig.~\ref{fig:halo_evo} is presented as a red curve in
Fig~\ref{fig:mah}. This halo stays in the rapid growth phase until
$z=1.2$, and its assembly is dominated by smooth spherical infall as
Fig.~\ref{fig:halo_evo} indicates. This appears quite different from
the hierarchical growth described by \citet{wechsler02} and
\citet{zhao03}: their rapid growth phase is dominated by mergers.
They also speculated that the universal inner density profile results
from violent relaxation during this fast merger phase. We will see
below that the density profile is also universal in our HDM
simulation. Thus the NFW profile apparently does not require
mergers, as noted originally by \citet{huss99}.

Halo formation histories obviously differ in CDM and HDM universes.
Another interesting question is how and where the first objects form
in the HDM case. Here we must keep in mind that the free-streaming
scale in HDM is just a characteristic scale and its relationship to
the actual mass of the first objects is not obvious. In
Fig.~\ref{fig:halo_evo2}, we present formation histories for two
``first'' objects. They were identified at $z=4.3$ (left column) and
$z=2.9$ (right column) when they had gathered $1215$ and $2686$
particles within $r_{200}$ respectively. In order to see the
particles better, we here use a lower  mass resolution simulation
glass128 which was also presented in Paper I. This simulation
has the same initial density field as HDMRUN, but 16 times lower
mass resolution. As in Fig~\ref{fig:halo_evo}, we trace back all
particles within $r_{200}$ (blue) and within $0.1r_{200}$ (red) to
higher redshift. Other background particles are shown as black
points. It is obvious that both haloes form by smooth spherical
collapse. In other words, the particles in the filaments and
``voids'' fall into the halo together. Even the particles within the
central region ($<0.1 r_{200}$) seem to follow a roughly spherical
collapse.  This agrees with what we see in Fig~\ref{fig:halo_evo}.
In the right column, the halo forms in a node which connects three
filaments, and the halo forms almost at the same time as the three
neighbour filaments. For the case in the left column, the halo forms
at the end of one filament ($z=6.0$) and then grows at the same time
as another filament. We have also checked some other examples and
find that almost all of them are born in nodes or at the ends of
filaments and grow by spherical accretion.

Merger events in such a monolithic universe are expected to be
rare. We find this to be true in our HDM simulations. For example,
in HDMRUN, if we define merger events as the merging of `real'
haloes whose masses are larger than the mass limit, only a few  
haloes experience a major merger event (with 
mass ratio of the two progenitors greater than $1:3$). 
We exclude these haloes from our study in this paper. 
Only about 30 percent of the remaining haloes experience  
minor merger events, and it is thus reasonable to regard them 
as a sample of `first' haloes.

\begin{figure*}
\bc
\hspace{-1cm}
\resizebox{16cm}{!}{\includegraphics{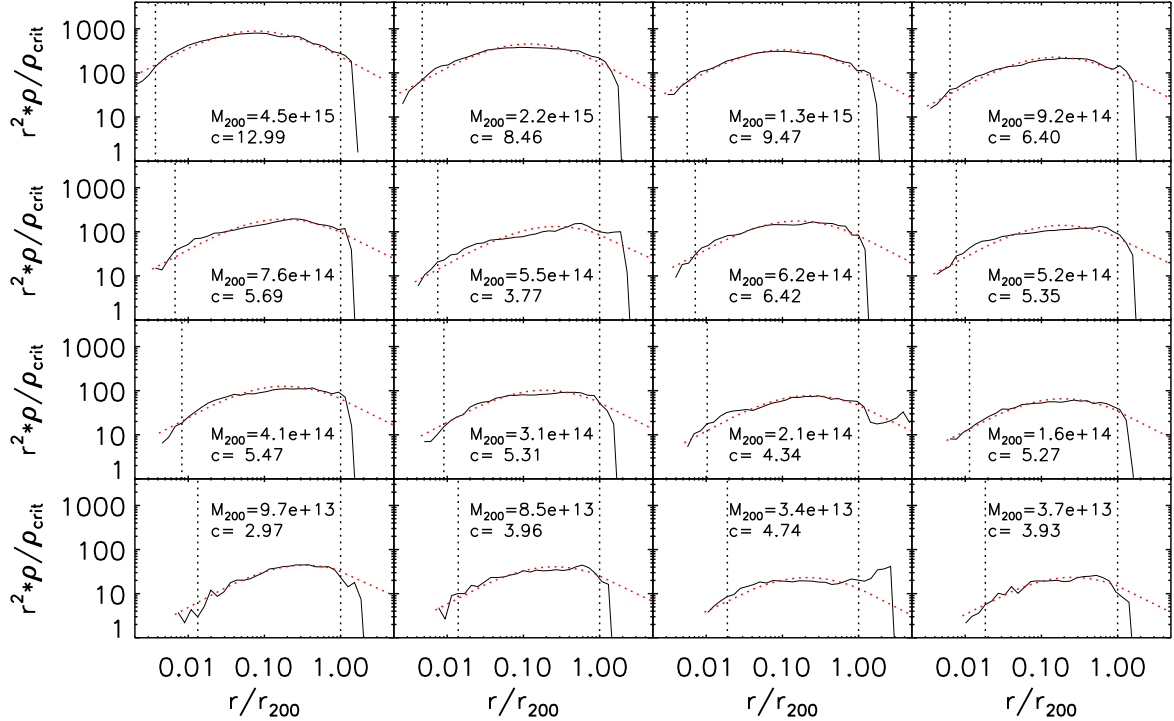}}\\%
\caption{Density profiles for 16 haloes with particle number within
$r_{200}$ ($N_{200}$) greater than 10,000. These 16 haloes cover
two orders of magnitude in mass. The density profiles are
normalized by $r^2/\rho_{crit}$ and the radius are normalized by
$r_{200}$. In each panel, the red dotted curve is the NFW fit to the
numerical measurement (black solid curve). Two vertical dotted lines
show the softening length and $r_{200}$. The corresponding $M_{200}$
(in unit of $h^{-1}M_{\odot}$ ) and the concentration
 parameter c are listed in each panel.}
\label{fig:halo_prof}
\ec
\end{figure*}
\begin{figure}
\bc
\hspace{-1.4cm}
\resizebox{8cm}{!}{\includegraphics{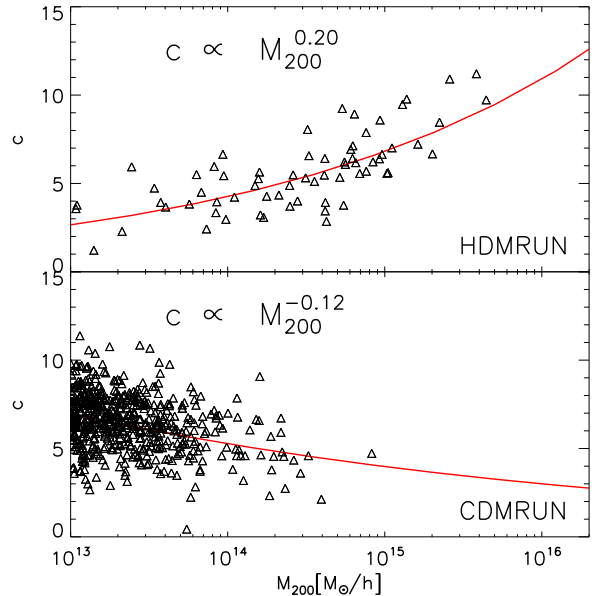}}\\%
\caption{The distributions of halo mass $M_{200}$ and concentration
parameter c. The top panel is from HDMRUN and the bottom
one from CDMRUN. }
\label{fig:c_m}
\ec
\end{figure}

\begin{figure}
\bc
\hspace{-1.4cm}
\resizebox{9cm}{!}{\includegraphics{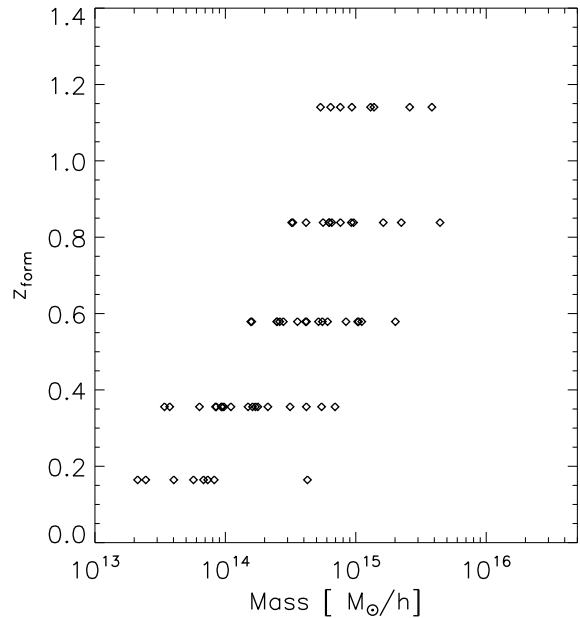}}\\%
\caption{The distribution of formation time for haloes with $M_{200} \geq
10^{13}M_{\odot}/h$  in the HDMRUN sample.}
\label{fig:t_m}
\ec
\end{figure}

\section{Halo properties in the two universes}
\label{sec:result}
In this section, we will study a range of ``first'' halo properties
in the HDM universe and compare them with their counterparts in
the $\Lambda$CDM model.

\subsection{Density Profile and Formation History}
Previous studies \citep{moore99a,colin00,eke01,busha07} have focused
on the formation of the first haloes in a WDM cosmology and have
claimed that the density profiles of these haloes do not differ
substantially from the NFW form found in a CDM universe. We support
these findings also in our HDM case, confirming the original result
of \citet{huss99b}. In Fig.~\ref{fig:halo_prof}, we present density
profiles for 16  haloes with mass from $3\times 10^{13} M_{\odot}/h$
to $4.5 \times 10^{15} M_{\odot}/h$. All these haloes include more than
$10,000$ particles.  Here the concentration parameter is measured by
fitting an NFW profile to the numerical results using
logarithmically spaced radial bins in the range $2\epsilon < r <
r_{200}$. The softening length $\epsilon$ and $r_{200}$ are shown in
the plot by vertical dotted lines. We find these profiles to follow
the NFW model very well. It is also obvious that the $\rho_s$ and
$c$ have a strong mass dependence. Both parameters increase with
increasing halo mass. These mass dependences disagree with those in
a CDM universe where more massive halos have a lower $\rho_s$ and
$c$.  In Fig.~\ref{fig:c_m}, we display the dependence of
concentration parameter on halo mass for our two halo samples. In
the lower panel, we present results for CDMRUN. The mass dependence,
$c \propto M_{200}^{-0.12}$, is close to that found by
\citet{neto07} and \citet{maccio07} $c \propto M_{200}^{-0.11}$ . 
In the upper panel, we present this
relation for HDMRUN. It is interesting that the mass dependence
inverts and follows $c \propto M_{200}^{0.2}$: the more massive a
halo, the larger its concentration parameter.

Many previous studies have found that the  structural properties and
the mass accretion histories of haloes are closely related in a CDM
universe \citep[e.g.][]{nfw96,nfw97,wechsler02,zhao03}.
Concentration increases with the formation time and the
characteristic density $\rho_s$ can be related with the mean cosmic
density at the time of formation. In Fig.~\ref{fig:t_m} , we check
this relation for our HDMRUN haloes. The formation time is defined
here as the earliest time when half of the halo mass was in its main
progenitor. It is interesting that massive haloes form a bit earlier
than their low mass counterparts in this cosmology. Combining with
the $M_{200} \sim c$ relation showed in Fig.~\ref{fig:c_m}, we find
that the earlier forming objects do indeed have a larger
concentration parameter, in agreement with the result for a CDM
universe. This indicates that here also the inner part is assembled
during the fast early growth phase.

\begin{figure}
\bc
\hspace{0cm}
\resizebox{9cm}{!}{\includegraphics{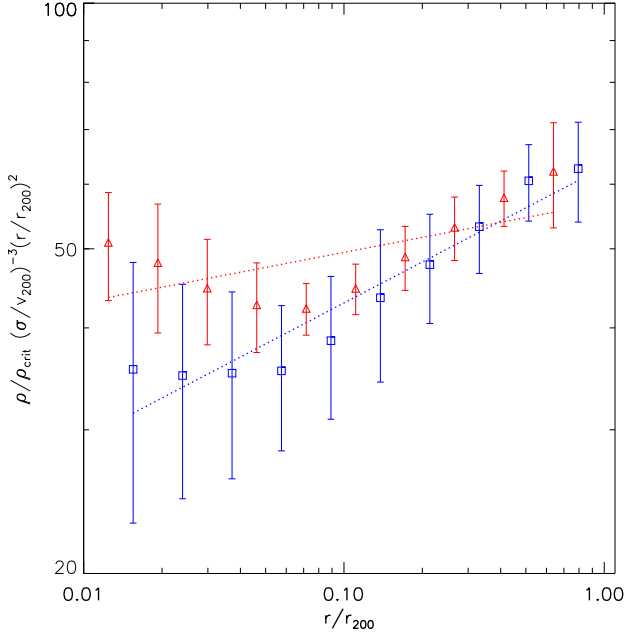}}\\%
\caption{Stacked phase density profiles for the 20 most massive
haloes in our two halo samples. The profile of each halo is
normalized by a factor $v_{200}^3/\rho_{crit}$ before stacking. The
dotted straight lines are our power law fits with indices
$\alpha=1.94$ and $a=1.83$ for HDMRUN(red triangles) and CDMRUN
(blue squares) respectively. In order to get a better dynamic range,
all symbols and lines multiplied by $(r/r_{200})^2$. The error-bars
indicate the $1\sigma$ scatter.} \label{fig:rho_phase} \ec
\end{figure}

\begin{figure}
\bc
\hspace{0cm}
\resizebox{9cm}{!}{\includegraphics{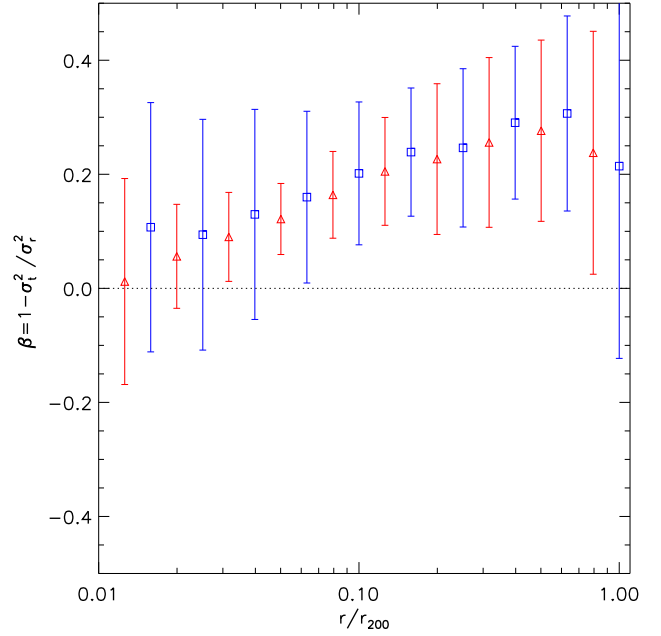}}\\%
\caption{The stacked velocity anisotropy profile $\beta(r)$ for the
20 most massive haloes in the HDMRUN (red triangles) and
CDMRUN (blue squares) samples. The error-bars indicate the
$1\sigma$ scatter.}
\label{fig:beta}
\ec
\end{figure}

\subsection{Kinematics}
If haloes follow a universal density profile as discussed above, and
are in an equilibrium state, then the solution of Jeans equation for
spherical, isotropic systems indicates that their kinematic
structure may also be universal. \citet{taylor01} have found that,
for dark haloes in a CDM universe, there is a power-law relationship
between ``phase-space density'' and radius: $\rho/\sigma^3 \approx
r^{-\alpha}$ with $\alpha=1.875$, where phase-space density is defined
as the ratio of local matter density $\rho$ to the cube of the local
velocity dispersion $\sigma$. This phase-space density  is inversely
related to the local entropy density. In the semi-analytic extended 
secondary infall model, 
this nearly scale-free nature of $\rho/\sigma^3$ is a robust feature
of virialized haloes in equilibrium \citep{austin05}. Further
investigation of halo formation processes indicates that this
scale-free feature cannot be the result of hierarchical merging;
rather it must be an outcome of violent relaxation \citep{austin05,barnes06}. 
We show results for massive haloes in
our HDM and CDM samples in Fig.~\ref{fig:rho_phase}. In order to
reduce the noise, we stack the 20 most massive haloes in each case.
Before the stacking, the profile for each halo is normalized by a
factor $V_{200}^3/\rho_{crit}$. The error-bars in the plot indicate
the $1\sigma$ scatter. We find the two profiles to follow a power
laws moderately well. The fitted indices  differ slightly:
$\alpha=1.94$ and $1.83$ for HDMRUN (red) and CDMRUN (blue)
respectively. 

In addition, the velocity anisotropy in haloes depends on
radius: The dispersion tensor is isotropic near the center and
moderately radially anisotropic near the virial radius. In
Fig.~\ref{fig:beta}, we present averaged velocity anisotropy
profiles $\beta(r)=1-\sigma_{t}^2/\sigma_{r}^2$ for the 20 most
massive haloes in our CDM and HDM samples. Here $\sigma_{t}$ and
$\sigma_{r}$ are the tangential and radial velocity dispersions
respectively. We again find very similar results for the two cases.
Outside $0.2r_{200}$, the mean radial anisotropy in HDMRUN (red) is
slightly larger than in CDMRUN (blue), but the effect is very small.

The similarities in phase-space density profile and velocity
anisotropy profile in the two cases indicate that these properties
are also universal and depend little on whether a halo is assembled
by mergers or by monolithic collapse.

\begin{figure}
\bc
\hspace{0cm}
\resizebox{9cm}{!}{\includegraphics{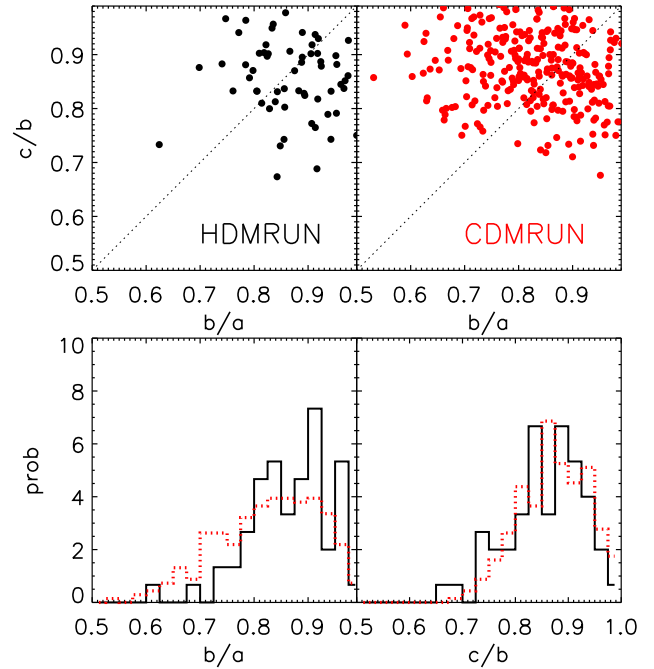}}\\%
\caption{Scatter plots of the axis ratios of haloes in HDMRUN and
CDMRUN (top two panels) and histograms of their probability
distributions (lower two panels: red dotted and black solid curves
are histograms for CDMRUN and HDMRUN respectively). Here a, b and c
are the major, intermediate and minor axis lengths.
Kolmogorov-Smirnov tests indicate that the distributions in two
universes do not differ significantly. } \label{fig:shape} \ec
\end{figure}

\subsection{Halo Shape}
In Fig.~\ref{fig:shape}, we show axis ratios for haloes in HDMRUN
and CDMRUN. We define the axes using the inertia tensor of the mass
distribution within the virial radius $r_{200}$:
\begin{equation}
I_{\alpha\beta}=\frac{1}{N_p}\Sigma_{i=1}^{N_p}r_{i,\alpha}r_{i,\beta}.
\end{equation}
where the $r_{i}$ are the positions of the $N_p$ particles within
$r_{200}$. $\alpha$ and $\beta$ are tensor indexes with values of
1,2 or 3 and indicate the three components of each particle's
position. After diagonalizing this matrix, characteristic axis
lengths are found as the square root of the eigenvalues. In the top
two panels of Fig.~\ref{fig:shape}, the intermediate-major ($b/a$) and 
minor-intermediate ($c/b$) axis ratios of halo samples from the two
simulations are presented. The distributions of shape parameters are
clearly very similar in the two cases. The mean values of the axis
ratios $b/a$ and $c/b$ are both close to $0.82$.  In the lower two
panels, we compare the probability distributions of $b/a$ and $c/b$
in the two simulations. The Kolmogorov-Smirnov test shows no
evidence for any difference between the distributions in the HDM and
CDM cases.

The above results indicate that different growth paths
(monolithic or hierarchical) affect the global shape of halos very
little. 

\begin{figure}
\bc
\hspace{-1.4cm}
\resizebox{8cm}{!}{\includegraphics{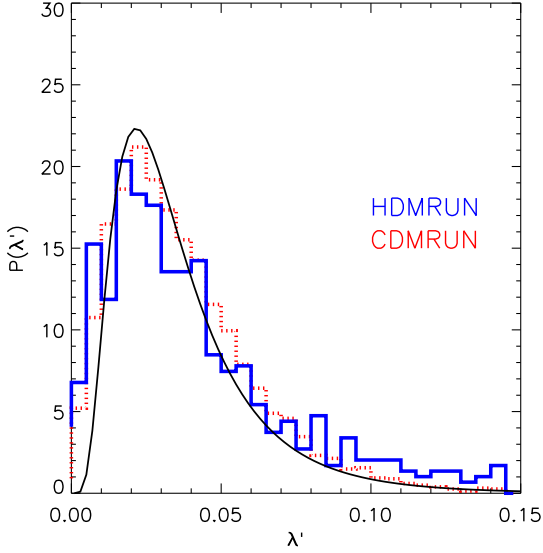}}\\%
\caption{The probability distribution of the spin parameter
$\lambda'$ in HDMRUN (blue solid) and CDMRUN (red dotted). The black
solid curve is the fit to a CDM universe from \citet{bullock01b}.}
\label{fig:spin} \ec
\end{figure}

\begin{figure}
\bc
\hspace{-1.4cm}
\resizebox{8cm}{!}{\includegraphics{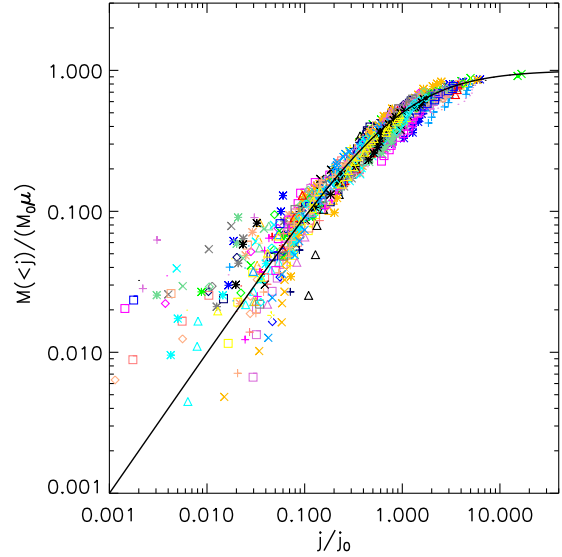}}\\%
\caption{Mass distributions of specific angular momentum within 58
massive haloes in HDMRUN. Each coloured symbol presents one halo.
Each mass distribution is normalized by its own virial mass ($\rm
M_0$) and shape parameter $\mu$, all symbols then stay around the
black curve $\frac{x}{1+x}$.} \label{fig:dist_ang} \ec
\end{figure}

\begin{figure}
\bc
\hspace{-1.4cm}
\resizebox{8cm}{!}{\includegraphics{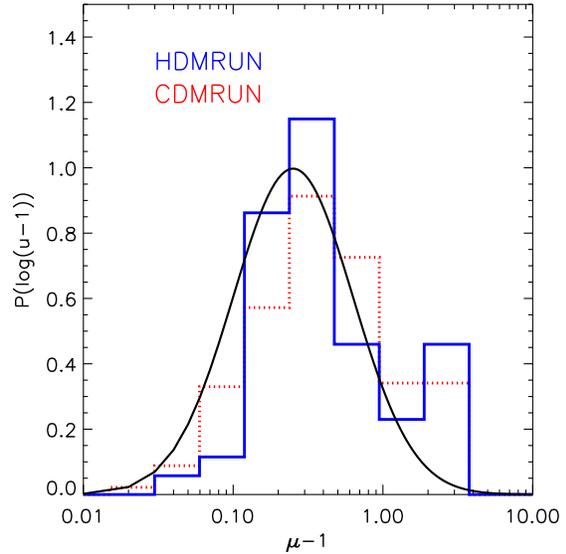}}\\%
\caption{The probability distribution of the shape parameter $\mu$
in HDMRUN (blue solid) and CDMRUN (red dotted). The black solid
curve is the fit to a CDM universe from \citet{bullock01b}.}
\label{fig:spin_u} \ec
\end{figure}

\subsection{Spin and Angular Momentum Distribution within Haloes}
The conventional measure of halo angular momentum, the dimensionless
spin parameter $\lambda$, is defined in terms of mass, energy and
angular momentum and is related to the ratio between a halo's mean
angular velocity ($\omega$) and the angular velocity which would be
required to support it by rotation alone($\omega_0$)
\citep{peebles69}:

\begin{equation}
\label{equ:spin1} \lambda \equiv \frac{J |E|^{1/2}}{GM^{5/2}} \simeq
0.4 \frac{\omega}{\omega_0}.
\end{equation}
The total angular momentum $J$ and energy $E$ are needed to
calculate this parameter. However, in this study, we follow
\citet{bullock01b} and define a more easily measured spin parameter
$\lambda'$ as:
\begin{equation}
\lambda'=\frac{J}{\sqrt{2}M_{200}Vr_{200}}
\end{equation}
where $J$ is the total angular momentum of all particles within
$r_{200}$ and  $V=\sqrt{GM_{200}/r_{200}}$ is the circular velocity
at radius $r_{200}$. This definition gives similar value to
Equ.~\ref{equ:spin1}. Because $\lambda'$ is defined for isolated
systems but applied to haloes in their cosmic context, the practical
definition of a halo is  more critical than the choice of definition
of $\lambda$ \citep{bett07}.

The nett spin of a dark halo is acquired from torques exerted by
neighbouring structures at early times \citep{hoyle49,peebles69,
doroshkevich70, white84}, and does not evolve much after the
turnaround point of a halo's MAH. After this time, the moment of
inertia of the collapsing material decreases and the universal
expansion reduces the strength of tidal forces
\citep{porciani02}. \citet{vitvitska02} noted that the spin
parameter fluctuates strongly with time, depending on the details of
assembly: the spin increases abruptly during a major merger and
decreases gradually between such mergers. \citet{donghia07} found
this effect to reflect the unrelaxed nature of the system;
equilibrium haloes show no significant correlation between spin and
merging history. In our HDM simulation, most haloes grow in a
monolithic way, so the impact from major mergers should be
negligible.

In Fig.~\ref{fig:spin}, we compare the distribution of spin
parameters in our HDMRUN halo sample with that in the CDMRUN sample.
A Kolmogorov-Smirnov test shoes no significant difference between
two different cases.

\citet{bullock01b} noticed that the distribution of specific angular
momentum $j$ within the CDM haloes has a universal profile,
specifically that the distribution of mass over specific angular
momentum $j$ is well fit in most haloes by the universal form:
\begin{equation}
M(<j)=M_{0}\frac{\mu j}{j_0+j}
\end{equation}
where $M_{0}$ is the virial mass, which we here replace by
$M_{200}$. $j_0=(\mu-1)j_{max}$ and $j_{max}$ is the maximum
specific angular momentum. This behaviour is not truly universal
since $\mu$ is an adjustable shape parameter which varies from halo
to halo. We now check this distribution for haloes in our HDM and
CDM simulations.

In order to resolve the $j$ profile adequately, we only consider
haloes with particle number larger than $3\times 10^4$. Different
regions are defined as cells in the usual spherical coordinates
$(r,\theta,\phi)$. We make sure that each cell contains an
approximately equal number of particles. We first divide the whole
halo into 10 radial shells which host almost equal particle numbers.
Then each shell is further divided into six zones: each cell spans
the full $2\pi$ range in $\phi$ and spans the equal solid angle
between $\rm cos(\theta)=-1$ and 1. The two zones with the same r
and  $\rm |cos(\theta)|$ that are above and below the equatorial
plane are assigned to one cell. In Fig.~\ref{fig:dist_ang}, we
present the distribution of normalized mass fraction $ M(<j)/(M_{0}
\mu)$ for 58 haloes from HDMRUN. Different coloured symbols present
different haloes.  We find that, after the parameter $\mu$ has been
adjusted, all haloes are close to the curve $x/(1+x)$, thus the
specific angular momentum profile can be described by one shape
parameter $\mu$ very well.  We compare the probability distribution
of this parameter for the HDMRUN and CDMRUN samples in Fig.
~\ref{fig:spin_u}. The distributions are similar and close to a
log-normal distribution as noted by \citet{bullock01b} ( A
Kolmogorov-Smirnov test shows no indication of a difference.) A
similar result was found by \citet{chen02} for WDM haloes: the
distribution is statistically indistinguishable from the CDM case.
As in \citet{vandenbosch02} and \citet{chen02}, we also find some
cells with negative angular momentum. These are excluded from our
analysis.

The similarity between the distributions of spin parameter and
internal specific angular momentum in the HDM and CDM cases shows
that formation through mergers is not necessary to generate
``universal'' angular momentum distributions. This conflicts with
the explanation which \citet{vitvitska02} give for the origin of the
angular  momentum distribution within CDM haloes.

\begin{figure}
\bc
\hspace{-1.4cm}
\resizebox{9cm}{!}{\includegraphics{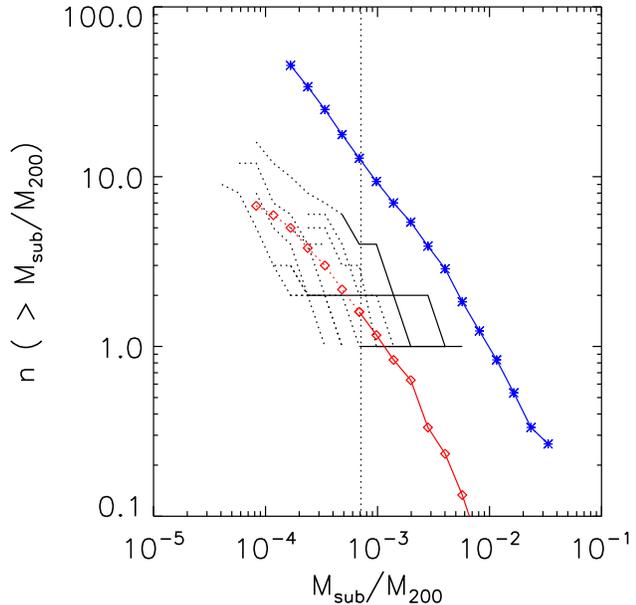}}\\%
\caption{The stacked cumulative subhalo mass function of the 30 most
massive haloes for our two samples: red diamonds and blue stars
are for HDM and CDM haloes respectively. The results for several
individual HDM haloes are also presented as black curves. The HDM
curves are divided into two parts at $0.1 M_{lim}$: real (spurious)
subhaloes have masses greater (less) than this mass limit and are
shown with solid (dotted) curves. The vertical dashed line indicates
the mean mass limit for all 30 haloes in HDMRUN $<0.1 M_{lim}>$.}
\label{fig:mf_sub} \ec
\end{figure}

\subsection{Substructure}
In Paper I, we gave a detailed discussion of the formation of
substructures in a HDM Universe. We found almost all
substructures in HDM haloes are either a result of spurious filament
fragmentation (at low masses) or major mergers at late times (at
large masses). As we see in Fig.\ref{fig:halo_evo}, at $z=0.84$ and
$z=0.58$, there are many artificial regularly spaced objects falling
into the clusters from filaments and these become substructures of
the HDM haloes. This effect prevents us from studying the abundance
of the real subhaloes at low mass since it is hard to identify which
subhaloes (if any) are real. In Fig~\ref{fig:mf_sub}, we present
cumulative mass functions of subhaloes for stacks of the 30 most
massive haloes in each sample. For the HDM case, the subhalo
mass functions of a few individual haloes are presented with black
curves. For haloes in the HDM universe, we define `real' subhaloes to
have mass greater than $0.1 M_{\rm lim}=8.8\times
10^{12}M_{\odot}/h$, and smaller subhaloes are considered
`spurious'. The mass function of the `real' and `spurious' subhaloes
are shown as solid and dotted lines respectively. The vertical
dashed line indicates the mean mass limit $<0.1 M_{lim}>$ for all 30
haloes in the HDMRUN. We have assumed here that the average mass
loss fraction is $0.1$ when a subhalo falls into its parent halo and
suffers tidal stripping. It is interesting that both
distributions follow a similar power law. That of the HDMRUN sample
is approximately one order of magnitude lower than the counterpart
in CDMRUN and on average very few `real' subhaloes are found in each
halo in the HDM case. This reflects the lack of hierarchical build
up in the HDM case.

\section{Conclusions and Discussion}
\label{sec:conclusions} In this paper , we study the haloes which
form by monolithic collapse in HDM cosmological simulations, and we
compare a range of properties of these haloes with these of their
counterparts in a $\Lambda$CDM hierarchical universe. From this
comparison we explore which physical mechanisms are responsible for
the universal profiles or distributions of halo properties. Our
simulated HDM universe has an inherent characteristic scale below
which the formation of small haloes is suppressed. In this universe,
the first generation of halo forms in a ``top-down'' way -- more
massive haloes form earlier by smooth accretion. Mergers occur in
significant numbers only when building up later halo generations. The
HDM cosmology thus provides a good example of monolithic halo growth
from Gaussian initial condition. Furthermore, because the HDM
characteristic scale is much more sharply imprinted than in the WDM
case, we  have a large and well defined monolithic halo sample. As a
result, we can explore  monolithic growth in a better controlled way
than previous work based on WDM universes. We summarise our
conclusions on monolithic halo formation  as follows:

(i) First generation haloes grow by smooth accretion. Such haloes
form from roughly spherical collapse of matter around the nodes or
ends of filaments. The formation histories of these monolithic
haloes are also characterised by two phases: fast initial and slow
later accretion, very similar to those found in a CDM universe. In
the fast growth phase, smooth infall is the main mechanism rather
than the mergers seen in the CDM case.

(ii) The density profiles of our monolithic haloes are well fit by
the NFW profile. The concentration parameter $c$ and the
characteristic density $\rho_s$ have strong dependences on halo
 mass: the more massive a halo, the larger  $c$ and $\rho_s$.
This is the inverse of the CDM dependence, but the
concentration-formation time relation is quite similar in the two
cases: earlier formed haloes tend to be more concentrated than their
later formed counterparts.

(iii) Phase-space density profiles and velocity dispersion
anisotropy profiles are very similar in our two cases. This
indicates that the kinematic structure of haloes is generated by
wide variety of dynamical collapse processes, not just by mergers.

(iv) The distribution of shape parameter is very similar in HDM and
CDM universes, as are the distributions of spin parameter
$\lambda'$ and of internal specific angular momentum.

(v) In our HDM simulations, most subhaloes result from the infall of
spurious small haloes which form in the filaments as a result of
numerical discreteness. The subhalo abundance is much lower than in
CDM haloes, and indeed there are very likely no subhaloes in first
generation HDM haloes.

The above results show that, except for substructure, the properties
of haloes formed by monolithic collapse are very similar to those of
haloes formed by hierarchical clustering in a CDM universe. In
particular, they have very similar profiles of density, phase-space
density and specific angular momentum, and very similar
distributions of axial ratios and spin parameter. These results
indicate that mergers are not responsible (or are  not required) for
the universal properties of CDM haloes. This agrees with results
from earlier work based on constrained simulations \citep{huss99,
macmillan06}. Attempts to explain these universal properties with
merger-driven models seem unlikely to be on the right track.

The results in this study could also find some applications in
realistic dark matter models, especially for the WDM cosmology. In
our HDM universe, first generation haloes form by smooth
near-spherical accretion with initial mass well below the
characteristic free-streaming mass. These haloes form simultaneously
with the filaments and sheets. This picture is different from that
in the familiar CDM universe. Our results suggest that the structure
of the first haloes, in a WDM universe (and thus, presumably, also of
their later descendants) will differ little from the structure of
CDM haloes, despite the differences in assembly history. This makes
it seem unlikely that changing to a WDM model will help resolve the
apparent problems of the CDM model on galactic scales.

\section*{Acknowledgments}
We thank Volker Springel for providing us with the simulation code
{\small GADGET}2 and with post-processing software. JW thanks Carlos
Frenk for his hospitality at ICC, Durham where the draft of this
paper was finished. JW also thanks Liang Gao and LiXin Li for useful 
comments. The simulations described in this paper were carried out on the 
Blade Centre cluster of the Computing Center of the Max-Planck-Society 
in Garching.

\label{lastpage}
\bibliographystyle{mn2e}
\bibliography{hdm2}
\bsp
\end{document}